\documentclass[prd,aps,showpacs,showkeys,twocolumn,10pt,nofootinbib,altaffilletter]{revtex4-2}
\usepackage{amssymb,amsmath,amsthm}
\usepackage{mathrsfs}
\usepackage{mathtools}
\usepackage[applemac]{inputenc}
\usepackage[thinc]{esdiff}

\pdfoutput=1

\usepackage{xcolor}
\usepackage{graphicx}
\usepackage{dcolumn}
\usepackage{bm}
\usepackage[pdftex]{hyperref}

\hypersetup{
	colorlinks=true,
	linkcolor=black,
	citecolor=black,
	filecolor=black,
	urlcolor=black,
}

\begin{document}
	
	\title{The little rip in classical and quantum $f(R)$ cosmology}
	
	\author{Teodor Borislavov Vasilev}
	\email{teodorbo@ucm.es}
	\affiliation{Departamento de F\'isica Te\'orica and IPARCOS, Universidad Complutense de Madrid, \\E-28040 Madrid, Spain.}
	
	\author{Mariam Bouhmadi-L\'opez}%
	\email{mariam.bouhmadi@ehu.eus}
	\affiliation{
		IKERBASQUE, Basque Foundation for Science, 48011, Bilbao, Spain
	}%
	\affiliation{%
		Department of Physics, University of the Basque Country, UPV/EHU, P.O. Box 644, 48080 Bilbao, Spain.
	}

	\author{Prado Mart\'in-Moruno}
	\email{pradomm@ucm.es}
	\affiliation{
		Departamento de F\'isica Te\'orica and  IPARCOS, Universidad Complutense de Madrid, \\E-28040 Madrid, Spain.
	}

	\date{\today}
	
	\begin{abstract}
		The little rip is a cosmological abrupt event predicted by some phantom dark energy models that could describe the future evolution of our Universe. This event can be interpreted as a big rip singularity delayed indefinitely, although in those models bounded structures will be destroyed in a finite cosmic time in the future. In this work, we analyse the little rip cosmology from a classical and quantum point of view within the scheme of alternative metric $f(R)$ theories of gravity. The quantum analysis is performed in the framework of $f(R)$ quantum geometrodynamics by means of the modified Wheeler-DeWitt equation. In this context, we show that the DeWitt criterion can be satisfied. Similar to what happens in general relativity, this result points towards the avoidance of the little rip in $f(R)$ quantum cosmology.
	\end{abstract}
	
	\maketitle
	
	\section{Introduction}
	
	Understanding the mechanisms involved in the accelerated expansion of our Universe has become one of the greatest milestones in modern cosmology. Since the moment this acceleration was first discovered \cite{Riess,Perlmutter}, a large number of models have been proposed to explain the origin of this phase. In general relativity (GR), the accelerated expansion is attributed to an exotic form of energy with  negative pressure that pushes the cosmos further into expansion. This extra content is designated as dark energy (DE). However, the nature of this energy is still unknown. Among the DE models that best fit the existing observational data is the standard $\Lambda$CDM, where the cosmological constant $\Lambda$ plays the role of DE. Nevertheless, the theoretically expected and the observed values for $\Lambda$ differ in several orders of magnitude (for a review on the topic, see Ref.~\cite{Lambdaproblems}) and, therefore, one can expect that this is just a useful effective model. Thus, various alternative models for DE describing the late-time cosmology without a cosmological constant have been proposed. Some examples are scalar fields in the form of quintessence \cite{Caldwell,Tsujikawa}, k-essence \cite{kessenceChiba:1999ka}, DE of phantom nature \cite{PhantomDE}, tachyonic matter \cite{TachyonGibbons:2002md,TachyonPadmanabhan:2002cp}, Chaplygin gas \cite{ChaplyginKamenshchik:2001cp,ChaplyginBento:2002ps} and holographic DE \cite{HoloLi:2004rb}, among others.
	On the other hand, it is also possible to describe the same  cosmological evolution, which in GR is attributed to DE, without the introduction of new material content. In this fashion, alternative theories of gravity have acquired a renewed interest, since they can provide a different framework for explaining the accelerated expansion of our Universe. Accordingly, the late-time cosmic expansion  can be understood as a consequence of suitable modifications to the underlining theory of gravity rather than due to some exotic matter content. Some examples of this approach are 	$f(R)$ theories of gravity \cite{DEinfRNojiri}, models with nonminimal coupling between curvature and
	matter
\cite{Bertolami:2007gv} (see also Refs.~\cite{Nojiri:2004bi, Allemandi:2005qs}), Gauss-Bonnet gravity \cite{GB1Dehghani:2004cf,GB2Nojiri:2005jg}, $f(R,\mathcal{T})$ gravity \cite{DEinfRTHarko:2011kv}, where $\mathcal{T}$ stands for the trace of the energy momentum tensor,  $f(T)$ modified teleparallel gravity \cite{DEinfTBengochea:2008gz}, being $T$ the torsion scalar, modified symmetric teleparallel $f(Q)$ theories of gravity \cite{Jimenez:2019ovq}, where Q denotes the nonmetricity scalar, and Horndeski theories \cite{Horndeski} (see also, e.g., Ref.~\cite{Kase:2018aps}), among others. For a review on the state of the art of DE cosmology see, for example, Refs.~\cite{Huterer:2017buf,reviewDE} and references therein. 
	
	From a practical point of view, whatever the origin of DE may be, it can be described effectively by an equation of state parameter $w$. This parameter represents the ratio between the pressure and energy density of DE, or the effective energy density and pressure in the case of alternative theories of gravity.  
	Latest cosmological data suggest that $w$  lies on a very narrow band around -1 \cite{Planck1,Planck3}. Therefore, the possibility of DE being of phantom nature ($w<-1$)  is not observationally excluded. On the contrary, this is even suggested by some data \cite{RisalitiLusso} and could be a solution to alleviate the $H_0$ tension (see, for example, Refs.~\cite{EDVH01,EDVH02}). However, considering phantom DE raises back ancient questions regarding the final fate of our Universe, since phantom DE is well known for possibly leading the cosmic expansion towards a future singularity. All bounded structures in the universe and  space-time itself might be ripped apart at a final big rip (BR) singularity \cite{BR}, where the Hubble rate, its cosmic time derivative and the scale factor diverge at a finite cosmic time. The cosmos could also reach an infinite expansion rate at a finite size of the observable universe, freezing its evolution at a big freeze singularity \cite{BF1BouhmadiLopez:2006fu, BF2BouhmadiLopez:2007qb}. Nonetheless, if $w<-1$ but $w\to-1$ sufficiently rapidly, then the occurrence of future singularities may be infinitely delayed in time \cite{LR}, in which case the singularity is called an abrupt cosmic event. Indeed, the little rip (LR) abrupt event is just a big rip that would take place at the infinite asymptotic future, although bounded structures will be destroyed in a finite time from the present, see Refs.~\cite{LR,LR2}. Another abrupt cosmic event appearing in phantom DE is the little sibling of the big rip (LSBR). This event is characterized by the divergence of the Hubble rate and the scale factor at an infinite future cosmic time, while  the cosmic time derivative of the Hubble rate remains finite \cite{LSBR}. For observational constraints on these riplike cosmic catastrophes see Refs.~\cite{CosmoConstraints2,CosmoConstraints}. (See also Refs.~\cite{sudden1,sudden2,sudden3} for other examples of cosmic singularities  and Refs.~\cite{EOSalpha3,Dabrowski:2014fha, SingularityClasification} for a detailed classification of DE singularities.) Nevertheless, it is commonly believed that quantum gravity effects may  smooth or avoid these (classical) cosmic catastrophes, see Refs.~\cite{Dabrowski:2006dd,Kamenshchik:2007zj} (see also Refs.~\cite{EOSalpha3,SingularityClasification,EOSalpha1,Elizalde:2004mq,Nojiri:2004ip,BouhmadiLopez:2009pu}).

	Previous works on quantum cosmology have shown that the aforementioned phantom riplike doomsdays, namely the BR, the LR and the LSBR, can be avoided due to quantum effects rising up as the universe approaches the classical singularity \cite{Dabrowski:2006dd,GRLR,Albarran:2015cda}. Since the background late-time cosmology can be equivalently described in the context of GR or  by alternative theories of gravity, it is natural to wonder whether these singularities are still avoided in the quantum realm for a different underlying theory of gravity. Following this line of thought, it has already been established that the BR and the LSBR can be avoided due to quantum effects in $f(R)$ cosmology \cite{fRQC,LSBRfR}. However, up to our knowledge, no results for the LR abrupt event in quantum $f(R)$ cosmology have been published so far. Thus, in this work we address the missing  quantum fate of the LR in metric $f(R)$ theories of gravity. To do so, we consider a group of $f(R)$ theories of gravity that predict the occurrence of the LR abrupt event at the classical level. (For previous works on the LR cosmology in alternative theories of gravity see Refs.~\cite{LR3fR,LRinfHG,LR4fR}.) The quantum analysis is performed in the framework of $f(R)$ quantum geometrodynamics, with the Wheeler-DeWitt equation \cite{DeWitt} being adapted to the $f(R)$ gravity case \cite{Vilenkin}. Consequently, we explore the possibility of avoiding the LR cosmic doomsday in $f(R)$ quantum cosmology.
	
	This paper is organized as follows. In Sec.~\ref{sec:LRinGR}, we review some basic results and observational constraints on the LR abrupt event in the framework of GR. In Sec.~\ref{sec:LRinfR}, we consider that the LR could take place when the description of gravity is that provided by alternative $f(R)$ theories of gravity. For that aim, we briefly introduce the reconstruction method for metric $f(R)$ gravity in Sec.~\ref{sec:recon}. Thereafter, in Sec.~\ref{sec:fR LR}, we apply the reconstruction method to the LR model reviewed in Sec.~\ref{sec:LRinGR}. Thus, we obtain the group of metric $f(R)$ theories of gravity predicting the LR abrupt cosmic event, recovering the $f(R)$ function first presented in Ref.~\cite{LRinfHG}. In Sec.~\ref{sec:Qcosmology}, we study the quantum fate of the LR predicted in one of the theories obtained in the past section. We summarize previous results on the formulation of quantum cosmology in the framework of general $f(R)$ quantum geometrodynamics in Sec.~\ref{sec:MWD}.  Next, in Secs.~\ref{sec:Q LR} and \ref{sec:BO}, we take different approaches to solve the modified Wheeler-DeWitt equation  and we analyse the avoidance of the singularity by means of the DeWitt criterion. We summarize our results in Sec.~\ref{sec:conlusions}. Finally, we discuss the validity of the approximations carried out to solve the modified Wheeler-DeWitt equation in Appendix \ref{appendix:BOvalidity}.

	\section{The little rip\label{sec:LRinGR}}
	
	Let us begin with a brief summary on the phenomenology of the LR abrupt event  in GR. 
	Throughout this work we limit ourselves to homogeneous and isotropic cosmological scenarios, which are described by the Friedmann-Lema\^itre-Robertson-Walker (FLRW) metric given by the line element 
	\begin{equation}\label{metric}
		ds^2=-dt^2+a(t)^2ds^2_3 ,
	\end{equation}
	where $a(t)$ stands for the scale factor, $ds_3^2$ represents the three-dimensional metric and we have used the geometric unit system $8\pi G=c=1$.
	Then, for the content of the universe represented by a perfect fluid with energy density $\rho$ and pressure $p$, the Einstein field equations reduce to the Friedmann and  Raychaudhuri equations
	\begin{align}\label{FE}
		\frac{\dot{a}^2}{a^2}&=H^2=\frac{1}{3}\rho-\frac{k}{a^2}   ,\\
		\frac{\ddot{a}}{a} &=\dot{H}+H^2=-\frac{1}{2}\left(p+\frac{\rho}{3}\right) \label{FE2}  ,
	\end{align}
	respectively, where the dot represents the derivative with respect to the cosmic time, $H$ denotes the Hubble rate and $k$ indicates the spatial curvature of the Universe (not fixed at this point). In the common interpretation of cosmological data within GR's framework, this perfect fluid is constituted with three different species: radiation, (dark and baryonic) matter (M) and DE. According to the latest observations  \cite{Planck1, Planck3} the current density parameters for  M and DE are $\Omega_{M,0}\sim 0.315$ and $\Omega_{DE,0}\sim0.685$, respectively, whereas radiation is negligible at the present time. Thus, DE is the dominant cosmic ingredient today. Furthermore, it will be even more dominant in the future since matter tends to dilute (faster).
	Thence, from a practical perspective, the contribution of  matter and radiation can be neglected when analysing the asymptotic future evolution of these cosmological models.
	Consequently, we consider that $p$ and $\rho$ in the above equations are those corresponding to DE.
	
	The LR abrupt event can be understood as a BR singularity that has been delayed indefinitely. This abrupt event was first discovered in Ref.~\cite{EOSalpha2table} (see also Refs.~\cite{EOSalpha3,Nojiri:2005sr}) and, thereafter, named as ``little rip'' in Ref.~\cite{LR}. (See also Ref.~\cite{BouhmadiLopez:2005gk} where the LR was found in the context of brane cosmology.) For the occurrence of this cosmic event,  DE  can be modelled by the following equation of state (EoS)
	\begin{equation}\label{eos0.5}
		p=-\rho-A\sqrt{\rho},
	\end{equation}
	being $A$ a positive parameter. From the conservation of the energy momentum tensor it follows that the energy density $\rho$ evolves with the scale factor as
	\begin{equation}\label{rho0.5}
		\rho=\rho_0\left[1+\frac{3A}{2\sqrt{\rho_0}}\ln\left(\frac{a}{a_0}\right)\right]^{2},
	\end{equation}
	being $\rho_0$ and $a_0$ the current values of the DE density and the scale factor, respectively. Then, combining  the Friedmann equation (\ref{FE}) and  Eq.~(\ref{rho0.5}), the time dependence of the scale factor reads
	\begin{align}\label{eq:a(t)}
		a(t)=&a_0 \exp\Bigg\lbrace
		\frac{2\sqrt{\rho_0}}{3A}\Bigg[\left(1+\frac{3A}{2\sqrt{\rho_0}}\ln\frac{a_\star}{a_0}\right)\nonumber\\
		&\times\exp\left(\frac{\sqrt{3}A}{2}\left(t-t_\star\right)\right)-1\Bigg]\Bigg\rbrace,
	\end{align}
	where we have denoted by $t_\star$ some arbitrary (future) moment in the expansion history of the Universe from which we can safely assume DE is the only content of the cosmos and $a_\star$ represents the corresponding scale factor. The EoS parameter $w$ for the DE follows from Eqs.~(\ref{eos0.5}) and (\ref{rho0.5}). That is
	\begin{equation}\label{wEOS alpha1/2}
		w=-1-\frac{A}{\sqrt{\rho_0}+\frac{3A}{2}\ln\left(\frac{a}{a_0}\right)}.
	\end{equation}
	Note that the preceding EoS parameter is clearly less than $-1$ (phantom DE) and asymptotically approaches $-1$  as the universe expands. Nevertheless, the behaviour of the universe is not that of a de Sitter model since the DE density (\ref{rho0.5}) and pressure are not constant. Furthermore, they even tend to explode with the superaccelerated expansion of the universe, as found in Ref.~\cite{EOSalpha2table} (see also Ref.~\cite{LR} and references therein).
	On the other hand, the corresponding Hubble rate and its cosmic time derivative are
	\begin{align}
		H(t)&=\sqrt{\frac{\rho_0}{3}}\left(1+\frac{3A}{2\sqrt{\rho_0}}\ln \frac{a_\star}{a_0}\right)\exp\left[\frac{\sqrt{3}}{2}A\left(t-t_\star\right)\right],\label{H(t)}\\
		\dot{H}(t)&=\frac{A}{2}\sqrt{\rho_0}\left(1+\frac{3A}{2\sqrt{\rho_0}}\ln \frac{a_\star}{a_0}\right)\exp\left[\frac{\sqrt{3}}{2}A\left(t-t_\star\right)\right]\label{dotH(t)}.
	\end{align}
	Therefore, in this model the scale factor $a$, the Hubble parameter $H$ and its cosmic derivative $\dot{H}$ diverge in the infinite distant future.  Nonetheless, bounded structures are shown to  be disintegrated in a finite time from present \cite{LR}. Moreover, disintegration can take place even before than for the BR scenario \cite{LR}.
	
	From an observational point of view, the DE model described by the EoS given in Eq.~(\ref{eos0.5}) has been shown to be compatible with the current expansion history of our Universe; see Ref.~\cite{CosmoConstraints}. In fact, authors in Ref.~\cite{CosmoConstraints} found a preferred value for the parameter $A$ when binding the model with current observational data.  Here we have denoted this value as $\bar{A}$. This is (in geometric units $8\pi G=c=1$)
	\begin{eqnarray}\label{Aconstrained}
		\bar{A}=2.75\times10^{-28} \ \textup{m}^{-1},
	\end{eqnarray}
	where m stands for metres. Note that the small value obtained there suggests that tiny deviations form the $\Lambda$CDM scenario are, indeed, the observational preferred situation \cite{CosmoConstraints}.

	Even though in the present work we will focus on the LR abrupt event modelled by the EoS given in Eq.~(\ref{eos0.5}), it should be stressed that this is not the only scenario where a fate \textit{\`a la} LR occurs. For example a more general EoS, still of the form of Eq.~(\ref{eos0.5}), was considered in Refs.~\cite{EOSalpha3,EOSalpha1}. This is
	\begin{equation}\label{EOS alpha}
		p=-\rho-A\rho^{\alpha}.
	\end{equation}
	This EoS was  thoroughly discussed in terms of singularity occurrence in Ref.~\cite{EOSalpha2table}. As there concluded, depending on the value of $\alpha$ several types of DE-driven singularities (at finite time from present epoch) may occur. 
	Nevertheless, if $\alpha\leq1/2$ the singularity is infinitely delayed in time; thus, an abrupt event takes place. In this case, a straightforward examination of the Hubble rate and its time derivative reveals that $\dot{H}$ remains finite and $H$ diverge when $\alpha\leq0$. This corresponds to a final fate \textit{\`a la} LSBR. 
	Whereas both quantities diverge for $0<\alpha\leq1/2$. Hence, for the latter range of values for $\alpha$, the DE described by EoS (\ref{EOS alpha}) leads the universe towards a LR abrupt event. 
	For other examples of LR cosmology see Refs.~\cite{LR,LR2} (see also Refs.~\cite{LR3fR,LRinfHG,LR4fR} for LR cosmology in $f(R)$ gravity).
	
	
	\section{The LR in $f(R)$ theories of gravity\label{sec:LRinfR}}
	
	Given some cosmological background evolution it is possible to find an alternative theory of gravity that leads to the same expansion history. The group of techniques used to perform such a ``reconstruction" task are commonly known as ``reconstruction methods" (for a review on the topic see, for example, Ref.~\cite{ReconNojiri99pag} and references therein).  In this section we focus on reconstruction methods within the framework of metric $f(R)$ alternative theories of gravity. Thence, we shall look for an $f(R)$ theory of gravity able to reproduce the superaccelerated expansion of the relativistic model filled with phantom DE described by the EoS (\ref{eos0.5}). Subsequently, as the LR abrupt event is inevitable in the latter case, then the reconstructed $f(R)$ theory will suffer the same classical fate.
	For previous works on reconstruction techniques in $f(R)$ gravity see, for instance, Refs.~\cite{ReconNojiri99pag,ReconCapozziello,ReconfRgravity,ReconMGravity,ReconNojiri,ReconDunsby,ReconCarloni}. See also Refs.~\cite{fRQC,LSBRfR,ReconRef1} for successful reconstruction of phantom DE-driven riplike events in metric $f(R)$ theories of gravity.
	
	
	\subsection{The reconstruction method\label{sec:recon}}

	Following a line of reasoning similar to that presented in Ref.~\cite{ReconRef1} (see also Ref.~\cite{LRinfHG}) we consider two cosmological evolutions to be equivalent at the background level if the corresponding geometrical variables $H$, $\dot{H}$, $R$ and $\dot{R}$ are identical. 
	
	Within GR, the expansion of the isotropic and homogeneous relativistic universe is ruled by the Friedmann and Raychaudhuri equations (\ref{FE}) and (\ref{FE2}), respectively. Accordingly, the scalar curvature reads
	\begin{equation}\label{R}
		R=6\left(\dot{H}+2H^2+\frac{k}{a^2}\right)=\rho-3p.
	\end{equation}
	From the continuity equation for the perfect fluid, i.e. 
	\begin{equation}
		\dot{\rho}+3H(p+\rho)=0,
	\end{equation}
	and the Friedmann equation (\ref{FE}), it follows
	\begin{align}\label{dot rho and p}
		\dot{\rho}&=-3(p+\rho)\left(\frac{1}{3}\rho-\frac{k}{a^2}\right)^{\frac{1}{2}}   , \\
		\dot{p}&=-3(p+\rho)\left(\frac{1}{3}\rho-\frac{k}{a^2}\right)^{\frac{1}{2}}\frac{dp}{d\rho}   ,
	\end{align}
	where we have assumed $p=p(\rho)$. Therefore, the cosmic time derivative of the scalar curvature $R$ is given by
	\begin{equation}\label{dot R}
		\dot{R}=-3\,(p+\rho)\left(\frac{1}{3}\rho-\frac{k}{a^2}\right)^{\frac{1}{2}}\left(1-3\frac{dp}{d\rho}\right)  .
	\end{equation}
	
	On the other hand, in the framework of metric $f(R)$ theories of gravity, the evolution of the universe is described by the action
	\begin{eqnarray}
		S=\frac{1}{2}\int d^4x\sqrt{-g}f(R)+S_m,
	\end{eqnarray}
	where $S_m$ stands for the minimally coupled matter fields.
	Under these circumstances, the field equations are no longer Eqs.~(\ref{FE}) and (\ref{FE2}). In fact the first field equation, the so-called modified Friedmann equation, reads
	\begin{equation}\label{M.F.E.}
		3H^2 \frac{df}{dR}=\frac{1}{2}\left(R\frac{df}{dR}-f\right)-3H\dot{R}\frac{d^2f}{dR^2	}-3\frac{k}{a^2}+\rho_m ,
	\end{equation}
	being $\rho_m$ the energy density of the minimally coupled matter fields. In the reconstruction method, the above expression is considered as a differential equation for some, \textit{a priori} unknown, function $f(R)$, where the coefficients are already fixed. That is, when the geometrical quantities involved in Eq.~(\ref{M.F.E.}) are set to be equal to those of the GR model that we want to reproduce, then the background cosmological expansion of the resulting metric $f(R)$ theory of gravity will be equivalent to that provided by the general relativistic model. Therefore, in the next section we solve Eq.~(\ref{M.F.E.}) for $f$ when considering that the behaviour of $H$, $R$ and $\dot{R}$ is given by the relativistic formulas (\ref{FE}) to (\ref{dot R}).
	As we are interested on the asymptotic behaviour of the universe, we can neglect the matter part, which will be quickly redshifted in the future. Furthermore, we shall focus our attention to flat FLRW for the same reason, as the contribution of the spatial curvature $k$ dilutes ($\propto a^{-2}$) with the cosmic expansion and, ultimately, will become unimportant when compared with the phantom DE density.
	
	
	\subsection{$f(R)$ theories predicting the LR\label{sec:fR LR}}
	
	We are mainly interested in $f(R)$ functions that mimics the behaviour given in Eq.~(\ref{eq:a(t)}), i.e. on $f(R)$ functions that describe the same asymptotic future behaviour of a FLRW universe to that of GR whose matter content corresponds to a DE fluid whose EoS reads as in Eq.~(\ref{eos0.5}). Under those ans\"atze, the matter content corresponding to baryonic and dark matter can be neglected.

	Note that the Hubble parameter in Eq.~(\ref{H(t)}) is an exponential function of the cosmic time. Thus, its time derivative is proportional to itself. For the sake of simplicity we denoted by $\beta$ that proportionality constant; i.e.
	\begin{equation}
		\dot{H}=\beta H,
	\end{equation}
	where we have defined
	\begin{eqnarray}
		\beta\coloneqq\frac{\sqrt{3}}{2}A.
	\end{eqnarray}
	Consequently, it is straightforward to express the curvature scalar and its time derivative in terms of the Hubble parameter. These are
	\begin{align}\label{R(H)}
		R&=6 H\left(\beta+2H\right),\\
		\dot{R}&=6\beta H\left(\beta +4H\right).\label{dotR(H)}
	\end{align}
	As a result of these expressions, Eq.~(\ref{M.F.E.}) simplifies when  rewritten in terms of the Hubble rate. Thus, the modified Friedmann equation, with $\rho_m=0$ and $k=0$, transforms into
	\begin{align}
		&\beta H^2\left(\beta+4H\right)f_{HH}-\left[4\beta H^2+H\left(\beta+H\right)\left(\beta+4H\right)\right]f_H\nonumber\\
		&+\left(\beta+4H\right)^2f=0.
	\end{align}
	The general solution to the preceding second-order differential equation was already found in Ref.~\cite{LRinfHG}, that is
	\begin{align}\label{sol:LRinf(R)}
		f(H)=&C_1\left(H^4-5\beta H^3+2\beta^2H^2+2\beta^3H\right)\nonumber\\
		&+C_2\bigg[\beta H\left(\beta^2+4\beta H-H^2\right)e^{\frac{H}{\beta}}\nonumber\\
		& +\left(H^4-5\beta H^3+2\beta^2H^2+2\beta^3H\right)\textup{Ei}\left(\frac{H}{\beta}\right)\bigg],
	\end{align}
	being $C_1$ and $C_2$ integration constants and Ei the exponential integral function (see definition 5.1.2 in Ref.~\cite{libroFunciones}). 
	
	On the other hand, to obtain the  expression for $f(R)$ we need to calculate the inverse of Eq.~(\ref{R(H)}). This is
	\begin{equation}\label{eq:H(R)}
		H=\frac{1}{12}\left(-3\beta\pm\sqrt{9\beta^2+12R}\right).
	\end{equation}
	However, since the scalar curvature increases along with the DE density, see (\ref{R})  and (\ref{rho0.5}), and the universe goes through an eternal expansion ($H>0$), only the positive branch of the solution is compatible with our model. Thus, we choose the positive sign in expression (\ref{eq:H(R)}). Therefore, we get \cite{LRinfHG}
	\begin{align}\label{f(R)}
		f(R)&=c_1\left[27\beta^4+150\beta^2R-\beta\left(9\beta^2+12R\right)^{\frac32}+2R^2\right]\nonumber\\
		&+c_2\Bigg\{\beta\left(-3\beta^2-2R+9\beta\sqrt{9\beta^2+12R}\right)\nonumber\\
		&\times\left(-3\beta+\sqrt{9\beta+12R}\right)\exp\left(-\frac14+\frac14\sqrt{1+\frac{4R}{3\beta}}\right)\nonumber\\
		&+ \left[27\beta^4+150\beta^2R-\beta\left(9\beta^2+12R\right)^{\frac32}+2R^2\right]\nonumber\\
		&\times\textup{Ei}\left(-\frac14+\frac14\sqrt{1+\frac{4R}{3\beta}}\right)\Bigg\},
	\end{align}
	where $c_1$ and $c_2$ are arbitrary constants. We emphasize that the metric $f(R)$ theory of gravity presented in Eq.~(\ref{f(R)}) leads to the same cosmological expansion as the DE fluid described by the EoS (\ref{eos0.5}) in the context of GR. However, in the framework of the modified theory of gravity, the evolution of the DE content is mimicked by the modifications appearing in the Friedmann equation as a result of $f(R)\neq R$. Consequently, since the relativistic model is doomed to evolve towards the LR abrupt event, then the reconstructed theory in Eq.~(\ref{f(R)}) will suffer the same classical fate. 
	We  further discuss on the viability of the theory given by Eq.~(\ref{f(R)}) in Sec.~\ref{sec:Q LR}.
	
	
	\section{The LR in quantum $f(R)$ cosmology\label{sec:Qcosmology}}
	
	The quantum fate of classical singularities can be addressed in the framework of quantum cosmology: the application of quantum theory to the Universe as a whole (see Refs.~\cite{KlausBarbara,Klauslibro} for a review on the topic). Although there are different approaches to quantum cosmology, we focus on one of the first attempts to quantize cosmological backgrounds \cite{DeWitt,Kuchar:1989tj}. This quantum cosmology is based on a canonical quantization with the Wheeler-DeWitt equation playing a central role \cite{DeWitt,Kuchar:1989tj,Wheeler}. In addition, following the ideas presented in Ref.~\cite{Vilenkin}, we adapt the Wheeler-DeWitt equation to $f(R)$ theories of gravity. The resulting scheme is known as $f(R)$ quantum geometrodynamics. Within this framework, we analyse the singularity avoidance by means of the DeWitt (DW) criterion \cite{DeWitt}. This is, the classical singularity might be avoided due to quantum effects if the wave function of the Universe vanishes in the configuration space close to the event. Therefore, this criterion is based on a probabilistic interpretation of the wave function. Note that this criterion has been successfully applied in several cosmological scenarios; see,  e.g., Refs.~\cite{SingularityClasification,BouhmadiLopez:2009pu,Dabrowski:2006dd,GRLR,Albarran:2015cda,fRQC,LSBRfR,Bouhmadi-Lopez:2016dcf,Bouhmadi-Lopez:2018tel,Albarran:2018mpg} among others.

	\subsection{Modified Wheeler-DeWitt equation\label{sec:MWD}}
	
	Cosmological models in $f(R)$ theories of gravity can be described, in the so-called Jordan frame, by the action
	\begin{equation}
		S=\frac{1}{2}\int d^4x \sqrt{-g}f(R).
	\end{equation}
	When considering a FLRW background geometry, the above action can be cast into 
	\begin{equation}\label{eqLagrangianDensityf(R)}
		S=\frac{1}{2}\int dt \ \mathcal{L}(a,\dot{a},\ddot{a}),
	\end{equation} 
	where the Lagrangian reads
	\begin{equation}
		\mathcal{L}(a,\dot{a},\ddot{a})=\mathcal{V}_{(3)}\,a^3f(R),
	\end{equation} 
	denoting by $\mathcal{V}_{(3)}$ the spatial three-dimensional volume. 
	It is well known that metric $f(R)$ theories of gravity carry an extra degree of freedom in comparison with GR (Einstein and Jordan formulation of $f(R)$ gravity can be found in \cite{EvsJframe1,EvsJrframe2,EvsJrframe3,EvsJrframe4} and references therein).  Therefore, for the canonical quantization of alternative $f(R)$ theories of gravity a new variable can be introduced to emphasize the existence of an extra degree of freedom. Furthermore, for a suitable choice of this new variable the second derivatives of the scale factor appearing in Eq.~(\ref{eqLagrangianDensityf(R)}) can be removed. Following the line of reasoning presented in Ref.~\cite{Vilenkin}, we select $R$ to be the new variable. Thus, the action in Eq.~(\ref{eqLagrangianDensityf(R)}) becomes
	\begin{eqnarray}
		S=\frac{1}{2}\int dt \ \mathcal{L}(a,\dot{a}, R, \dot{R}).
	\end{eqnarray}
	Nevertheless, since  $R$ and $a$ are not independent
	(at the classical level), their relation needs to be  properly introduced in the theory via a Lagrange multiplier, $\mu$, for the constraint $R=R(a,\dot{a},\ddot{a})$. Thence,
	\begin{equation}
		\mathcal{L}=\mathcal{V}_{(3)} a^3\left\lbrace f(R)-\mu\left[ R-6\left(\frac{\ddot{a}}{a}+\frac{\dot{a}^2}{a^2}+\frac{k}{a^2}\right) \right]\right\rbrace   .
	\end{equation}
	The Lagrange multiplier can be determined varying the action with respect $R$. This is
	\begin{eqnarray}
		\mu=f_R(R),
	\end{eqnarray}
	with the notation $f_R\coloneqq df/dR$.
	Then, the Lagrangian can be reformulated as
	\begin{align}\label{L}
		\mathcal{L}(a,\dot{a},R,\dot{R})=&\mathcal{V}_{(3)}\Big\{ a^3\Big[f(R)-R f_R(R)\Big]\nonumber\\
		&-6a^2f_{RR}(R)\dot{a}\dot{R}+6af_R(R)(k-\dot{a}^2)\Big\},
	\end{align}
	where $f_{RR}\coloneqq d^2f/dR^2$. Thereafter, in order to ease the application of the quantization procedure, we proceed to diagonalize the derivative part by the introduction of a new set of variables like in Ref.~\cite{Vilenkin}. These are
	\begin{subequations}\label{Vilenkin}
		\begin{align}
			\label{Vilenkin q} 
			q&\coloneqq a\sqrt{R_\star}\left(\frac{f_R}{f_{R_\star}}\right)^{\frac{1}{2}},\\
			x&\coloneqq\frac{1}{2}\ln\left(\frac{f_R}{f_{R_\star}}\right),\label{Vilenkin x} 
		\end{align}
	\end{subequations}
	being $R_\star$ a constant (not fixed at this point) needed for the above change of variables to be well defined. 
	In the new variables the Lagrangian (\ref{L}) transforms into
	\begin{align}\label{L(x,q)}
		\mathcal{L}(x,\dot{x},q,\dot{q})=&\mathcal{V}_{(3)}\left(\frac{R_\star f_R}{f_{R_\star}}\right)^{-\frac{3}{2}}q^3 \left[f-6f_R\frac{\dot{q}^2}{q^2}\right.  \nonumber\\
		&\left. -Rf_R +6f_R\dot{x}^2+6k\frac{R_\star}{f_{R_\star}}\frac{f^2_R}{q^2}\right],
	\end{align}
	where $f$ and $f_R$ are now understood as functions of $x$. This form of the Lagrangian is already suitable for the quantization procedure.
	
	Since the kinetic part has been diagonalized, the derivation of the  Hamiltonian is straightforward. The conjugate momenta read
	\begin{align}
		P_q&=\frac{\partial\mathcal{L}}{\partial \dot{q}}=-12\mathcal{V}_{(3)} R_\star^{-\frac{3}{2}}f^{\frac{3}{2}}_{R_\star}f^{-\frac{1}{2}}_R q \dot{q} ,\\
		P_x&=\frac{\partial\mathcal{L}}{\partial \dot{x}}=12\mathcal{V}_{(3)} R_\star^{-\frac{3}{2}}f^{\frac{3}{2}}_{R_\star}f^{-\frac{1}{2}}_R q^3 \dot{x}.
	\end{align}
	Then, the corresponding Hamiltonian is
	\begin{align}
		\mathcal{H}=&-\mathcal{V}_{(3)}q^3\left(\frac{R_\star f_R}{f_{R_\star}}\right)^{-3/2} \bigg\{ f+6k\frac{R_\star}{f_{R_\star}}\frac{f_R^2}{q^2} \nonumber\\
		&-Rf_R+\frac{6R_\star^3}{(12)^2\mathcal{V}_{(3)}^2f_{R_\star}^3}\frac{f_R^2}{q^4}\left[P_q^2-\frac{P_x^2}{q^2}\right]\bigg\}.
	\end{align}
	
	For the canonical quantization procedure, we assume $P_q\to-i\hbar\partial_q$ and $P_x\to -i\hbar\partial_x$. As a result, the classical Hamiltonian constraint $\mathcal{H}=0$ becomes the modified Wheeler-DeWitt (mWDW) equation for the wave function $\Psi$ of the Universe \cite{DeWitt,Vilenkin,Klauslibro}. Therefore, the mWDW equation reads
	\begin{equation}
		\mathcal{\hat{H}}\Psi=0.
	\end{equation}
	After suitable rearrangements, the preceding expression can be cast in the form of the hyperbolic differential equation \cite{Vilenkin}
	\begin{equation}\label{MWDW}
		\Big[\hbar^2q^2\partial^2_q-\hbar^2\partial^2_x-V(q,x)\Big]\Psi(q,x)=0 ,
	\end{equation}
	where the effective potential is given by
	\begin{equation}\label{V}
		V(q,x)=\frac{q^4}{\lambda^2}\left(k+\frac{q^2}{6R_\star f_{R_\star}}(f-Rf_R)e^{-4x}\right)  ,
	\end{equation}
	with $\lambda\coloneqq  R_\star/(12\mathcal{V}_{(3)}f_{R_\star})$. Note that when the expression of the $f(R)$ is given, the variables $x$ and $q$   are univocally fixed. Then, $f$ and $Rf_R$ must be rewritten in terms of $x$. In the next section, we will again focus on the case $k=0$.
	

	\subsection{Asymptotic wave function close to the LR\label{sec:Q LR}}
	
	From here on, we consider the expression for $f(R)$ gravity found in Eq.~(\ref{f(R)}) by means of the reconstruction methods applied in Sec.~\ref{sec:fR LR}. We recall that the group of metric $f(R)$ theories of gravity presented in Eq.~(\ref{f(R)}) lead to the same cosmological evolution as the relativistic model filled with phantom DE described by the EoS (\ref{eos0.5}). Hence, both models share a common final fate: the LR abrupt event.
	
	It is important to note that, for a given $f(R)$ expression, the analytic inversion of the definition for the variable $x$ in Eq.~(\ref{Vilenkin x}) is critical for the computation of the mWDW equation. This is because the relation $R=R(x)$ is crucial in order to express the effective potential (\ref{V}) in terms of $x$. 
	Unfortunately, the term multiplying $c_2$ in Eq.~(\ref{f(R)}) prevent us from inverting relation (\ref{Vilenkin x}) due to the presence of the exponential integral function Ei. Therefore, for the sake of simplicity, we consider $c_2=0$. Thus, we study the subgroup of metric $f(R)$ theories of gravity with a LR abrupt event given by
	\begin{align}\label{chosenf(R)}
		f(R)=c_1\left[27\beta^4+150\beta^2R-\beta\left(9\beta^2+12R\right)^{\frac32}+2R^2\right].
	\end{align}
	For this particular $f(R)$, the change of variables (\ref{Vilenkin}) reads
	\begin{align}
		q&=a\sqrt{\frac{2c_1R_\star}{f_{R_\star}}}\left(75\beta^2-9\beta\sqrt{9\beta^2+12R}+2R\right)^\frac12,\\
		x&=\frac12\ln\left[\frac{2c_1}{f_{R_\star}}\left(75\beta^2-9\beta\sqrt{9\beta^2+12R}+2R\right)\right],\label{Vilenkin x LRinfR}
	\end{align}
	with $f_{R_\star}=2c_1\left(75\beta^2-9\beta\sqrt{9\beta^2+12R_\star}+2R_\star\right)$. 
	Concerning the value of $R_\star$,  note that Eqs.~(\ref{eos0.5}), (\ref{rho0.5}) and (\ref{R}) imply
	\begin{align}\label{R(a)}
		R=&4\rho_0\left[1+\beta\sqrt{\frac{3}{\rho_0}}\ln\left( \frac{a}{a_0}\right)\right]^2\nonumber\\
		&+
		6\beta\sqrt{\frac{\rho_0}{3}}\left[1+\beta\sqrt{\frac{3}{\rho_0}}\ln \left(\frac{a}{a_0}\right)\right].
	\end{align}
	Thus, following the spirit for a physical meaningful $R_\star$ discussed in Ref.~\cite{Vilenkin} (and also in Refs.~\cite{fRQC,LSBRfR}), we define this constant as the value of the scalar curvature evaluated at some future scale factor $a=a_\star$ on which the description of the universe by means of DE only becomes appropriate. For the sake of concreteness, we set this moment to occur at $a_\star=100a_0$ \footnote{At that point in the expansion, the matter content will be diluted with respect to the present concentration by a factor of $100^{3}$. Roughly speaking this is $\Omega_M=10^{-6}\Omega_{M,0} \approx 3.06\times10^{-7}$ and, therefore, $\Omega_{DE}\approx 1$.}. Thence,
	\begin{align}\label{Rstar}
		R_\star\coloneqq&4\rho_0\left(1+\beta\sqrt{\frac{3}{\rho_0}}\ln100\right)^2\nonumber\\
		&+
		6\beta\sqrt{\frac{\rho_0}{3}}\left(1+\beta\sqrt{\frac{3}{\rho_0}}\ln100\right).
	\end{align} 
	As already discussed, $A$ and $\rho_0$ are constrained by observations [see Eq.~(\ref{Aconstrained}) (and Ref.~\cite{CosmoConstraints})], and, therefore, it can be seen that this definition for $R_\star$ makes the change of variables in Eq.~(\ref{Vilenkin}) well defined. Moreover, $f_R>0$ and $f_{RR}>0$ for all $R>R_\star$ when $c_1>0$ in Eq.~(\ref{chosenf(R)}), thus preventing the effective Newton's constant to become negative; see implications in Ref.~\cite{viablefR} and references therein.
	
	
	
	Next in the quantization procedure, we compute the inverse of Eq.~(\ref{Vilenkin x LRinfR}). This is
	\begin{align}
		R(x)=&84\beta^2+\frac{1}{4c_1}f_{R_\star}e^{2x}\nonumber\\
		&\pm54\beta\sqrt{\frac{1}{48c_1}f_{R_\star}e^{2x}+2\beta^2}.
	\end{align}
	However, only the positive branch is compatible with $R$ being an increasing function of $x$ and $R>R_\star$. Therefore, we choose the positive sign in the preceding expression.
	Then, the effective potential in the mWDW equation reduces to
	\begin{eqnarray}\label{VLR}
		V(q,x)=-U(x)q^6 ,
	\end{eqnarray}
	where
	\begin{widetext}
		\begin{align}\label{U}
			U(x)=&\frac{f_{R_\star}}{48c_1\lambda^2R_\star}\Bigg\lbrace1+1644\frac{c_1\beta^2}{f_{R_\star }}e^{-2x}+205992\frac{c_1^2\beta^4}{f_{R_\star }^2}e^{-4x}\nonumber \\
			&+36\beta e^{-x} \left(1+336\frac{c_1\beta^2}{f_{R_\star }}e^{-2x}\right)\sqrt{\frac{3c_1}{f_{R_\star}}\left(1+96\frac{c_1\beta^2}{f_{R_\star }}e^{-2x}\right)}\nonumber\\
			&-12\beta\sqrt{\frac{3c_1}{f_{R_\star}}} e^{-x}\left[1+330\frac{c_1\beta^2}{f_{R_\star} }e^{-2x}+18\beta e^{-x}\sqrt{\frac{3c_1}{f_{R_\star}}\left(1+96\frac{c_1\beta^2}{f_{R_\star }}e^{-2x}\right)}\right]\nonumber\\
			&\times\left[1+339\frac{c_1\beta^2}{f_{R_\star}}e^{-2x}+18\beta e^{-x}\sqrt{\frac{3c_1}{f_{R_\star}}\left(1+96\frac{c_1\beta^2}{f_{R_\star }}e^{-2x}\right)}\right]^\frac12\Bigg\rbrace.
		\end{align}
	\end{widetext}
	Since the main motivation of the present work is to evaluate the wave function $\Psi$ at the LR abrupt event, it is not strictly necessary to solve the  mWDW equation for the whole potential but focus only in the configuration space near the abrupt event. For that aim, note that the most important condition for the occurrence of the LR is the divergence of the scalar curvature. This corresponds to $q\to\infty$ and $x\to\infty$.
	Then, without loss of generality, we can suppose the variable $x$ to be large (but finite) when close to the LR. This assumption allows us to expand the above expression as
	\begin{align}\label{eq:U Asymp}
		U(x)\approx& \frac{f_{R_\star}}{48c_1\lambda^2R_\star}\left[1+24\beta\sqrt{\frac{3c_1}{f_{R_\star}}}e^{-x}+672\beta^2\frac{c_1}{f_{R_\star}}e^{-2x}\right.\nonumber\\
		&\left.+1152\beta^3\left(\frac{3c_1}{f_{R_\star}}\right)^\frac32 e^{-3x}+\mathcal{O}\left(e^{-4x}\right)\right].
	\end{align}
	Additionally, further approximations can be implemented. As the LR is approached, the majority of the terms appearing in the latter expansion are exponentially suppressed (since $x\to\infty$). Then, in order to analyse the asymptotic behaviour of $\Psi$, we keep only the dominant part of the effective potential\footnote{This simplification is also supported by the observational constraints on the model \cite{CosmoConstraints}. As the subdominant terms are proportional to some power of the parameter $A$ (remember $\beta=\sqrt{3}A/2$), then they are naturally suppressed since $A$ takes small values when observationally constrained, see expression (\ref{Aconstrained}).}. Therefore, at first-order approximation, we consider 
	\begin{equation}
		V(q,x)\approx -Bq^6,
	\end{equation}
	being $B\coloneqq\frac{f_{R_\star}}{48c_1\lambda^2R_\star}$. Consequently, the mWDW equation simplifies to
	\begin{equation}\label{leadingMWDW}
		\Big[\hbar^2q^2\partial^2_q-\hbar^2\partial^2_x+Bq^6\Big]\tilde{\Psi}(x,q)=0. 
	\end{equation}
	We have used the notation $\tilde{\Psi}$ to emphasize that this is the asymptotic form of Eq.~(\ref{MWDW}), where only the leading-order part of the effective potential is taken into account.
	This differential equation can be solved via a separation ansatz for the wave function of the Universe,
	\begin{eqnarray}\label{ansatz:Separation}
		\tilde{\Psi}(q,x)=\sum_{\tilde{k}}b_{\tilde{k}} \zeta_{\tilde{k}}(q)\theta_{\tilde{k}}(x),
	\end{eqnarray}
	being $b_{\tilde{k}}$  the amplitude of each solution and $\tilde{ k}$ standing for an integration constant related to the associated energy. Please, do not confuse $\tilde{k}$ with the spatial curvature $k$, which has been neglected. 
	As a result of the separation ansatz, the mWDW equation (\ref{leadingMWDW}) implies
	\begin{align}
		\hbar^2\diff[2]{\theta_{\tilde{k}}}{x} -{\tilde{k}}^2\theta_{\tilde{k}}&=0  \label{eq:theta} , \\
		\label{eq:zeta}
		\hbar^2q^2\diff[2]{\zeta_{\tilde{k}}}{q}+ \left(Bq^6-\tilde{k}^2 \right)\zeta_{\tilde{k}}&=0 .
	\end{align}
	The former equation can be straightforwardly solved and leads to
	\begin{align}
		\theta_{\tilde{k}}(x)=d_1 \exp\left(\frac{\sqrt{\tilde{k}^2}}{\hbar}x\right)+d_2\exp\left(-\frac{\sqrt{\tilde{k}^2}}{\hbar}x\right),
	\end{align}
	being $d_1$ and $d_2$  arbitrary constants. The above solutions correspond to trigonometric or exponential functions of $x$, depending on the sign of $\tilde{k}^2$.
	On the other hand,  the equation for  $\zeta_{\tilde{k}}$  can be solved in an exact way by means of Bessel functions; cf. 9.1.53 of Ref.~\cite{libroFunciones}. The solution can be written as
	\begin{align}\label{sol:zeta}
		\zeta_{\tilde{ k}}(q)=& \sqrt{q}\left[u_1 J_{\frac16\sqrt{1+4\frac{\tilde{ k}^2}{\hbar^2}}}\left(\frac{\sqrt{B}}{3\hbar}q^3\right)\right.\nonumber\\
		&\left.+u_2 Y_{\frac16\sqrt{1+4\frac{\tilde{ k}^2}{\hbar^2}}}\left(\frac{\sqrt{B}}{3\hbar}q^3\right)\right],
	\end{align}
	being $J$ and $Y$ the Bessel functions of first and second order, respectively, and $u_1$ and $u_2$ integration constants.
	
	Therefore, as $q$ and $x$ explode when approaching the LR, the solutions for the $\theta_{\tilde{k}}$ part remain finite if  the constant $d_1$ is set to zero for the case of $\tilde{ k}^2$ being positive. Whereas  the solutions for $\zeta_{\tilde{ k}}$  vanish as $q$ diverges since, for large values of $q$, Eq.~(\ref{sol:zeta}) reduces to
	\begin{align}
		\zeta_{\tilde{ k}}(q)\approx& \sqrt{\frac{6\hbar}{\pi}}\frac{1}{B^\frac14q}\left[\tilde{u}_1\exp\left(i\frac{\sqrt{B}}{3\hbar}q^3\right)\right.\nonumber\\
		&\left.+\tilde{u}_2 \exp\left(-i\frac{\sqrt{B}}{3\hbar}q^3\right)\right],
	\end{align}
	where $\tilde{u}_1$ and $\tilde{u}_2$ depend on $\tilde{k}$; cf. 9.2.1-2 of Ref.~\cite{libroFunciones}. Thus, the total wave function $\tilde{\Psi}$,
	\begin{align}\label{sol:mWDW LO Psi}
		\tilde{\Psi}(q,x)\approx& \sqrt{\frac{6\hbar}{\pi}}\frac{1}{B^\frac14q}\sum_{\tilde{k}}b_{\tilde{k}}\left[\tilde{u}_1\exp\left(i\frac{\sqrt{B}}{3\hbar}q^3\right)\right.\nonumber\\
		&\left.+\tilde{u}_2 \exp\left(-i\frac{\sqrt{B}}{3\hbar}q^3\right)\right]\left[d_1 \exp\left(\frac{\sqrt{\tilde{k}^2}}{\hbar}x\right)\right.\nonumber\\
		&\left.+d_2\exp\left(-\frac{\sqrt{\tilde{k}^2}}{\hbar}x\right)\right],
	\end{align}
	shrinks to zero as the LR abrupt event is approached. Hence we conclude that the DW condition is satisfied if one of the integration constants is fixed to zero.  Thus, as it happens in GR, this result hints towards the avoidance of the LR doomsday in $f(R)$ cosmology due to quantum gravity effects.

	Qualitatively, the wave function (\ref{sol:mWDW LO Psi}) has the same asymptotic form as exhibited in Ref.~\cite{LSBRfR} for the wave function of the LSBR abrupt event in $f(R)$ gravity. 
	However, owing to the fact that both events are different at the classical level, differences were expected to manifest in the shape of the wave functions. In fact, since the effective potentials entering the mWDW equation in each case are different, see  Eq.~(\ref{VLR}) and the analogous expression for the LSBR scenario given in Eq. (38) of Ref.~\cite{LSBR},
	then the similarity between both solutions seems to be an artefact of the approximations performed in order to solve the mWDW equations in the asymptotic limit. Hence, this resemblance is no longer expected to hold when
	further terms in the expansion (\ref{eq:U Asymp}) are taken into account.
	Following this reasoning, in the next section we propose a less restrictive approach that allows us to solve the mWDW equation for a wider region in the configuration space.

	
	\subsection{Born-Oppenheimer approximation for $\Psi$\label{sec:BO}}
	
	In this section, we address the solution of the complete mWDW equation,
	\begin{eqnarray}\label{NLOmWDW}
		\Big[\hbar^2q^2\partial^2_q-\hbar^2\partial^2_x+U(x)q^6\Big]\Psi(q,x)=0,
	\end{eqnarray}
	where  $U(x)$ is given by Eq.~(\ref{U}). 
	%
	Owing to the fact that the potential term in Eq.~(\ref{NLOmWDW}), that is $U(x)q^6$, now contains both variables, a separation ansatz like (\ref{ansatz:Separation}) will no longer apply. Instead, we propose an adiabatic semiseparability-type ansatz for the wave function of the Universe. This is based on the so-called Born-Oppenheimer (BO) ansatz, originally formulated in the context of molecular physics \cite{BOoriginal}. In cosmological scenarios, this approximation is often implemented such that the geometrical part of the total wave function (usually depending on the scale factor $a$) is factored out from the section encompassing the physical fields contained in the universe, which in turn are considered to depend adiabatically on the background geometry; see Refs.~\cite{ClausBO1,BOenWDW,ClausBO2} (for a recent work see, e.g., Ref.~\cite{BouhmadiLopez:2009pu}). In this way, it may feel tempting to apply the BO quasiseparability by factorizing the wave function $\Psi$ into a part depending only on $q$ and another related to both $q$ and $x$, since the scale factor $a$ enters only in $q$ and, therefore, disregarding the variable $x$ as carrying exclusively ``matter'' degrees of freedom. However, this way of reasoning is misleading. In fact, this naive separation results ultimately in both parts of the wave function depending on both original variables, $a$ and $R$ (since $a$ and $R$ enter the variable $q$). Consequently, following that procedure the interaction between $a$ and $R$ will be present in both parts of the wave function. Thus, this would make the quasiseparability ansatz pointless from the very beginning.
	The misconception leading to such failure is the innocent consideration that the mWDW is carrying two degrees of freedom of different nature; those are one geometrical and another rather related to matter fields. Although this is often true in cosmological scenarios, it is not longer the case for the mWDW equation of $f(R)$ cosmology, that is, Eq.~(\ref{MWDW}). Here we have two genuinely geometrical variables. These are the scale factor $a$ and the scalar curvature $R$, both contained in the definition of $q$ and $x$. Hence, a different formulation of the BO ansatz for solving Eq.~(\ref{NLOmWDW}) is needed.
	
	For that purpose, it should be stressed that $R$ can be considered to be more fundamental from a geometrical point  of view than the scale factor. Hence, if we are to conserve the spirit of the original BO ansatz, this is to quantize the geometry at first place and, after that, the remnant physical fields taking  into account the backreaction effects, then we should factorize $\Psi$ into a part depending only on $R$ (as the main geometrical variable) and another depending on $R$ and $a$. Therefore we propose the following ansatz \textit{\`a la} Born-Oppenheimer:
	\begin{eqnarray}\label{ansatz:nsTBO}
		\Psi(q,x)=\sum_{\tilde{k}}b_{\tilde{ k}}\chi_{\tilde{ k}}(q,x)\varphi_{\tilde{ k}}(x).
	\end{eqnarray}
	We emphasise that $x$ depends only on $R$, whereas $q$ contains both $a$ and $R$; see definitions in (\ref{Vilenkin}). In addition, $b_{\tilde{k}}$ stands for the amplitude of each solution and $\tilde{ k}$ is related with the associated energy.
	As a result, the mWDW equation (\ref{NLOmWDW}) reads
	\begin{align}\label{eq:twistedBO}
		&\hbar^2q^2\varphi_{\tilde{ k}}\diffp[2]{\chi_{\tilde{ k}}}{q}-\hbar^2\varphi_{\tilde{ k}}\diffp[2]{\chi_{\tilde{ k}}}{x}-2\hbar^2\diffp{\chi_{\tilde{ k}}}{x}\diff{\varphi_{\tilde{ k}}}{x}\nonumber\\
		&-\hbar^2\chi_{\tilde{ k}}\diff[2]{\varphi_{\tilde{ k}}}{x}+U(x)q^6\chi_{\tilde{ k}}\varphi_{\tilde{ k}}=0.
	\end{align}
	Then, the contribution of the second and third terms can be neglected due to the adiabatic assumption. (The validity of this approximations is justified in the Appendix \ref{appendix:BOvalidity}.) Thus, Eq.~(\ref{eq:twistedBO}) implies the following equations:
	\begin{align}
		\hbar^2\diff[2]{\varphi_{\tilde{ k}}}{x}-\tilde{ k}^2\varphi_{\tilde{ k}}&=0,\label{eq:NLO STBO varphi}\\
		\hbar^2q^2\diffp[2]{\chi_{\tilde{ k}}}{q}+\left[U(x)q^6-\tilde{ k}^2\right]\chi_{\tilde{ k}}&=0.\label{eq:NLO STBO chi}
	\end{align}
	The former equation can be  solved in the same fashion as  Eq.~(\ref{eq:theta}). The solutions are  exponential and trigonometric functions, depending on the sign of $\tilde{ k}^2$, 
	\begin{align}\label{sol:varphi}
		\varphi_{\tilde{ k}}(x)=d_1\exp\left(\frac{\sqrt{\tilde{ k}^2}}{\hbar}x\right)+d_2\exp\left(-\frac{\sqrt{\tilde{ k}^2}}{\hbar}x\right), 
	\end{align}
	being $d_1$ and $d_2$ integration constants.
	On the other hand, due to the adiabatic approximation, the potential term $U(x)$ appearing in Eq.~(\ref{eq:NLO STBO chi}) is treated like a (quasi)constant parameter when solving for $\chi_{\tilde{k}}$. Note that this assumption is supported on the fact that $U(x)$, given in Eq.~(\ref{U}), converges very quickly to a constant value when $\beta$ is observationally constrained [see the expansion in Eq.~(\ref{eq:U Asymp})].
	Thence, the most general solution for $\chi_{\tilde{k}}$ is, cf. 9.1.53 of Ref.~\cite{libroFunciones},
	\begin{align}\label{sol:chi}
		\chi_{\tilde{ k}}(q,x)=& \sqrt{q}\left[u_1 J_{\frac{1}{6}\sqrt{1+\frac{4\tilde{ k}^2}{\hbar^2}}}\left(\frac{\sqrt{U(x)}}{3\hbar}q^3\right)\right.\nonumber\\
		&\left.+u_2 Y_{\frac16\sqrt{1+\frac{4\tilde{ k}^2}{\hbar^2}}}\left(\frac{\sqrt{U(x)}}{3\hbar}q^3\right)\right],
	\end{align}
	with $u_1$ and $u_2$ integration constants.  
	
	Near the LR abrupt event, expression (\ref{sol:varphi}) remains bounded for $\tilde{ k}^2$ negative. However, when $\tilde{ k}^2$ is positive, $\varphi_{\tilde{ k}}$ is finite if and only if the constant $d_1$ is set to vanish. In contrast, the solutions for the $\chi_{\tilde{ k}}$ function have all the same asymptotic form. This is
	\begin{align}
		\chi_{\tilde{ k}}(q,x)\approx&\sqrt{\frac{6\hbar}{\pi}}\frac{1}{U(x)^{\frac14}q}\left[\tilde{u}_1\exp\left(i\frac{\sqrt{U(x)}}{3\hbar}q^3\right)\right.\nonumber\\
		&\left.+\tilde{u}_2\exp\left(-i\frac{\sqrt{U(x)}}{3\hbar}q^3\right)\right],
	\end{align}
	for large values of $q$,
	cf. 9.2.1-2 of Ref.~\cite{libroFunciones}, where the integration constants $\tilde{u}_1$ and $\tilde{u}_2$ now depend on $\tilde{k}^2$.  Therefore, the asymptotic form of the total wave function  $\Psi$ reads
	\begin{align}\label{sol:mWDW NLO Psi}
		\Psi(q,x)\approx& \sqrt{\frac{6\hbar}{\pi}}\frac{1}{U(x)^{\frac14}q}\sum_{\tilde{k}}b_{\tilde{k}}\left[\tilde{u}_1\exp\left(i\frac{\sqrt{U(x)}}{3\hbar}q^3\right)\right.\nonumber\\
		&\left.+\tilde{u}_2\exp \left(-i\frac{\sqrt{U(x)}}{3\hbar}q^3\right)\right]\nonumber\\
		&\times\left[d_1\exp\left(\frac{\sqrt{\tilde{ k}^2}}{\hbar}x\right)+d_2\exp\left(-\frac{\sqrt{\tilde{ k}^2}}{\hbar}x\right)\right].
	\end{align}
	As $U(x)$ tends to a constant value when $x$ explodes, the wave function cancels at the LR abrupt event when one of the integrations constants is set to zero, $d_1=0$ for $\tilde{k}^2$ positive, in accordance with the results of the previous section, thus pointing towards the avoidance of this fatal fate. Nevertheless, since  $U(x)>B$ and $U(x)\to B$ asymptotically, then the rate at which the wave function shrinks is increased with respect to the asymptotic approach performed in the previous section. Ergo, subdominant order contributions to the effective potential speed up the vanishing rate of the wave function $\Psi$. Furthermore, since the approximation presented in this section is less restrictive than the asymptotic approach previously performed, the wave function here obtained is valid in a broader region in the configuration space.

	
	\section{Conclusions\label{sec:conlusions}}
	
	The LR abrupt event is a cosmic  doomsday arising in some cosmological models where the accelerated expansion of the universe is driven by a DE of phantom nature. Since some of these models have been shown to be able to describe the current  cosmological observations \cite{CosmoConstraints},  our own Universe may evolve towards this singular fate. However, quantum effects can ultimately become significant and prevent the occurrence of such a doomsday. In fact, for the case of the background evolution being  that provided by GR, it has already been established that the DW criterion for singularity avoidance can be satisfied for the particular LR model considered here. Subsequently, in this work we have addressed the question whether this is still true when the classical evolution of the universe is due to an $f(R)$ metric theory of gravity.
	
	Hence, we have applied the so-called reconstruction methods to find a group of $f(R)$ theories of gravity that produce the same expansion history as that of a relativistic model filled with a DE fluid described by the EoS given in Eq.~(\ref{eos0.5}), thus obtaining the group of metric $f(R)$ theories of gravity that predict a classical fate \textit{\`a la} little rip.
	
	Thereafter, we have studied the quantum fate of the cosmos governed by one of the metric $f(R)$ gravity theories obtained before. The quantum analysis was performed within the framework of $f(R)$ quantum geometrodynamics, with the mWDW equation playing a central role. We have solved the mWDW equation and showed that the solutions satisfy the DW criterion when one of the integration constants is set to zero. Thus, as it also happens in general relativity, the fulfilment of the aforementioned condition hints towards the avoidance of this doomsday in $f(R)$ cosmology.
	
	Furthermore, we have performed different approaches when solving the mWDW equation. On a first approximation, we have solved the asymptotic form of the mWDW when only leading-order terms in the potential are considered.
	Afterwards, in a second approach, we have used a BO-type approximation (\ref{ansatz:nsTBO}) in order to analyse the behaviour of the wave function $\Psi$ for the complete mWDW equation. As a consequence of this less restrictive approach, the  resulting wave function exhibits a richer behaviour. Indeed, as the compliance region of the BO approach was shown to be wider than just the asymptotic regime, this method for solving the mWDW equation can be useful for comparing  the wave functions of different events. Especially when those events share a similar asymptotic regime, since the asymptotic approach to $\Psi$ will not find any differences there.
	
	It should be noted, however, that we have fixed to zero an integration constant in order to find vanishing solutions at the abrupt event. Therefore, we have disregarded a subgroup of solutions to the mWDW equation as unphysical. If future investigations show the importance of the dismissed solutions, then it would be concluded that the DW criterion may not always be fulfilled for solutions of physical interest.

	
	\section*{Acknowledgments}
	The research of T.~B.~V. and P.~M.~M. is supported by MINECO (Spain) Project No. PID2019-107394GB-I00 (AEI/FEDER, UE).
	T.~B.~V. also acknowledge financial support from Project No. FIS2016-78859-P (AEI/FEDER, UE) through Grant No. PAII46/20-08/2020-03, and from Universidad Complutense de Madrid and Banco de Santander through Grant No. CT63/19-CT64/19.
	The research of M.~B.~L. is supported by the Basque Foundation of Science Ikerbasque. She also would like to
	acknowledge the partial support from the Basque government Grant No. IT956-16 (Spain) and Project No. FIS2017-85076-P (MINECO/AEI/FEDER, UE).
	%
	
	
	
	\appendix
	
	\section{Validity of the BO approximation\label{appendix:BOvalidity}}
	
	During the application of the BO-type ansatz (\ref{ansatz:nsTBO}) performed in Sec.~\ref{sec:BO}, we have considered that  $\chi_{\tilde{ k}}(q,x)$ depends adiabatically on $x$. Therefore, we have neglected the contribution of some parts in Eq.~(\ref{eq:twistedBO}). This approach is valid as long as the corresponding solutions satisfy
	\begin{align}
		\hbar^2\varphi_{\tilde{ k}}\diffp[2]{\chi_{\tilde{ k}}}{x},& \ 2\hbar^2\diffp{\chi_{\tilde{ k}}}{ x}\diff{\varphi_{\tilde{ k}}}{x}\ll\nonumber\\
		& \hbar^2q^2\varphi_{\tilde{ k}}\diffp[2]{\chi_{\tilde{ k}}}{q},\ \hbar^2\chi_{\tilde{ k}}\diff[2]{\varphi_{\tilde{ k}}}{x},\ U(x)q^6\chi_{\tilde{ k}}\varphi_{\tilde{ k}}.
	\end{align}
	As a result of this approximation, the solutions for $\varphi_{\tilde{ k}}$ and $\chi_{\tilde{k}}$ are presented in Eqs.~(\ref{sol:varphi}) and (\ref{sol:chi}), respectively. Then, the terms we keep in (\ref{eq:twistedBO}) read
	\begin{align}
		&\hbar^2q^2\varphi_{\tilde{k}}\diffp[2]{\chi_{\tilde{k}}}{q}\approx-U(x)q^6\chi_{\tilde{k}}\varphi_{\tilde{k}}\approx-\sqrt{\frac{6\hbar}{\pi}}U(x)^\frac34q^5\nonumber\\
		&\times\left[\tilde{u}_1\exp \left(i\frac{\sqrt{U(x)}}{3\hbar}q^3\right)+\tilde{u}_2\exp \left(-i\frac{\sqrt{U(x)}}{3\hbar}q^3\right)\right]\nonumber\\
		&\times\left[d_1\exp\left(\frac{\sqrt{\tilde{ k}^2}}{\hbar}x\right)\right.\left.+d_2\exp\left(-\frac{\sqrt{\tilde{ k}^2}}{\hbar}x\right)\right],\\
		&\hbar^2\chi_{\tilde{k}}\diff[2]{\varphi_{\tilde{k}}}{x}\approx\sqrt{\frac{6\hbar}{\pi}}\frac{\tilde{k}^2}{U(x)^{\frac14}q}\nonumber\\
		&\times\left[\tilde{u}_1\exp \left(i\frac{\sqrt{U(x)}}{3\hbar}q^3\right)+\tilde{u}_2\exp \left(-i\frac{\sqrt{U(x)}}{3\hbar}q^3\right)\right]\nonumber\\
		&\times\left[d_1\exp\left(\frac{\sqrt{\tilde{ k}^2}}{\hbar}x\right)+d_2\exp\left(-\frac{\sqrt{\tilde{ k}^2}}{\hbar}x\right)\right].
	\end{align}
	While, the neglected terms behave asymptotically as
	\begin{align}
		&\hbar^2\varphi_{\tilde{k}}\diffp[2]{\chi_{\tilde{k}}}{x}\approx-\frac{1}{36}\sqrt{\frac{6\hbar}{\pi}}\frac{U'(x)^2}{U(x)^{\frac54}}q^5 \nonumber\\
		&\times\left[\tilde{u}_1\exp \left(i\frac{\sqrt{U(x)}}{3\hbar}q^3\right)+\tilde{u}_2\exp \left(-i\frac{\sqrt{U(x)}}{3\hbar}q^3\right)\right]\nonumber\\
		&\times\left[d_1\exp\left(\frac{\sqrt{\tilde{ k}^2}}{\hbar}x\right)+d_2\exp\left(-\frac{\sqrt{\tilde{ k}^2}}{\hbar}x\right)\right],
	\end{align}
	\begin{align}
		&2\hbar^2\diffp{\chi_{\tilde{k}}}{x}\diff{\varphi_{\tilde{k}}}{x}\approx\frac{i}{3}\sqrt{\frac{6\hbar\tilde{ k}^2}{\pi}}\frac{U'(x)}{U(x)^{\frac34}}q^2\nonumber\\
		&\times\left[\tilde{u}_1\exp \left(i\frac{\sqrt{U(x)}}{3\hbar}q^3\right)-\tilde{u}_2\exp \left(-i\frac{\sqrt{U(x)}}{3\hbar}q^3\right)\right]\nonumber\\
		&\times\left[d_1\exp\left(\frac{\sqrt{\tilde{ k}^2}}{\hbar}x\right)-d_2\exp\left(-\frac{\sqrt{\tilde{ k}^2}}{\hbar}x\right)\right].
	\end{align}
	Note that, for  $\tilde{ k}^2$ positive, the constants $d_1$ must be zero in order to have a vanishing wave function at the LR.
	Thus, to analyse the validity of the performed approximation we compare the largest of the neglected terms with the smallest of the saved ones. This is the ratio $\varepsilon$,
	\begin{align}\label{eqVal a/2}
		\varepsilon=\left|\frac{\hbar^2\varphi_{\tilde{ k}}\partial^2_x\chi_{\tilde{ k}}}{\hbar^2\chi_{\tilde{ k}}\partial^2_x\varphi_{\tilde{ k}}}\right|\approx\frac{U'(x)^2}{U(x)}\frac{q^6}{36\tilde{ k}^2}.
	\end{align}
	Consequently, the approximation is valid as long as $\varepsilon\ll1$. To obtain the compliance region of this condition, note 
	\begin{align}
		\frac{U'(x)^2}{U(x)}\approx&36\frac{\beta^2}{\lambda^2R_\star}e^{-2x}\left[1+\frac{40}{3}\sqrt{\frac{3c_1}{f_{R_\star}}}\beta e^{-x}\right.\nonumber\\
		&\left.+\frac{832}{3}\frac{c_1\beta^2}{f_{R_\star}}e^{-2x}+1152\sqrt{\frac{3c_1^3}{f_{R_\star}^3}}\beta^3e^{-3x}\right.\nonumber\\
		&+\mathcal{O}\left(e^{-4x}\right)\Bigg],
	\end{align}
	when $x$ is large. Then, in the configuration space near the cosmic event,
	\begin{eqnarray}
		\varepsilon\approx\frac{\beta^2}{\lambda^2\tilde{ k}^2R_\star}e^{-2x}q^6.
	\end{eqnarray}
	Finally, $\varepsilon\ll1$ near the LR  if $\beta$ is sufficiently small, i.e. for small value of $A$. Note that this corresponds, in fact,  to the observationally preferred situation \cite{CosmoConstraints}. [We recall that $A$ is of order $10^{-28}$ when observationally constrained, see Eq.~(\ref{Aconstrained})]. 
	%
	Therefore, when the parameters of the theory are observationally constrained, the approximation is valid throughout the semiclassical regime towards the abrupt event, where the variables $q$ and $x$ increase but not sufficiently rapidly to compensate the small value of $\beta^2$. Hence, for the purpose of this work, that is to analyse the fulfilment of the DW criterion in the configuration space close to the LR, this approximation is valid.
	
	\vfill\null

\end{document}